\documentclass[a4,article,nofootinbib,twocolumn]{revtex4-1}

\usepackage{graphicx}
\usepackage{dcolumn}
\usepackage{bm}
\usepackage{color}
\usepackage{amsmath}%
\usepackage{amsfonts}%
\usepackage{amssymb}%
\usepackage{breqn}
\usepackage{fixfoot}

\newtheorem{THM1}{\textit{Theorem}}[section]
\newtheorem{THM2}{\textit{Theorem}}[section]

\makeatletter
\let\cat@comma@active\@empty
\makeatother

\begin{document}
\title{A Stochastic Model of Inward Diffusion in Magnetospheric Plasmas}
\author{N. Sato and Z. Yoshida}
\affiliation{Graduate School of Frontier Sciences, The University of Tokyo,
Kashiwa, Chiba 277-8561, Japan}
\date{\today}

\begin{abstract}
The \emph{inward diffusion} of particles, often observed in magnetospheric plasmas
(either naturally created stellar ones or laboratory devices)
creates a spontaneous density gradient, which seemingly contradicts the entropy principle.
We construct a theoretical model of diffusion that can explain the inward diffusion in a dipole magnetic field. 
The key is the identification of the proper coordinates on which an appropriate diffusion operator can be formulated.
The effective phase space is foliated by the adiabatic invariants;
on the symplectic leaf, the invariant measure (by which the entropy must be calculated) is distorted,
by the inhomogeneous magnetic field, with respect to the conventional Lebesgue measure of the natural phase space.
The collision operator is formulated to be consistent to the ergodic hypothesis on the symplectic leaf,
i.e., the resultant diffusion must diminish gradients on the proper coordinates.
The non-orthogonality of the cotangent vectors of the configuration space
causes a coupling between the perpendicular and parallel diffusions,
which is derived by applying Ito's formula of changing variables.
The model has been examined by numerical simulations. 
We observe the creation of a peaked density profile that mimics radiation belts in planetary magnetospheres as well as laboratory experiments.

\end{abstract}

\maketitle

\section{Introduction}
Spontaneous creation of plasma clump is commonly observed in the vicinity of various-scale dipole magnetic fields,
ranging from planetary magnetospheres\,\cite{Voy1,Voy2,CRPWSE} to laboratory plasma devices\,\cite{Yos4, Box}.
However, if one were biased by the textbook knowledge that plasma particles are ``diamagnetic'', 
it is somewhat mysterious how a magnetic dipole can attract charged particles.
Or, if one calculates the Boltzmann distribution of charged particles, one finds that
the density does not respond to magnetic field.
We have to develop a reasonable model to explain how the \emph{inward diffusion}
(or, \emph{up-hill diffusion}) can take place to create a density gradient in a dipole magnetic field
---this is a theoretical challenge because the creation of gradient is seemingly contradicting the entropy principle.

The early works on the topic, mainly theoretical, date back to the $1960$s \cite{Falt,Bir}. 
The concept of electromagnetic fluctuations driven radial diffusion was investigated,
and a coarse-grained kinetic equation in magnetic coordinates was derived by a perturbation method. 
The key idea is the use of the scale separation, i.e., the time scale over which
the first and second adiabatic invariants (magnetic moment and bounce action respectively) are conserved is much longer than the characteristic time scale destroying the third adiabatic invariant (magnetic flux).

On the other hand, some more empirical models of radial diffusion were developed to explain actual observations in the Earth's magnetosphere: in \cite{Spj,Corn} the radial diffusion parameter is evaluated
by assuming an $\nu^{-2}$ spectrum, with $\nu$ the frequency of the electromagnetic fluctuations.
The main conclusion was that radial diffusion plays a central role in determining the stationary density profile.
This diffusion parameter was used in more recent works \cite{Fok2,Zheng} that deal with a bounce-averaged kinetic equation (see also \cite{Fok1,Fok3}) describing the electron radiation belts environment of our planet. 

Stimulated by the Voyager $1$ and $2$ missions  to Saturn \cite{Voy1,Voy2} and, more recently, by the Cassini Radio and Plasma Wave Science experiment \cite{CRPWSE},
intensive studies were made to explain the parallel pressure profile along magnetic surfaces ($L$-shells)
by the stationary force balance  \cite{Per1,Per2,Per3,Rich1,Rich2}.
As explained in \cite{Rich1}, the underlying assumption is that transport in the inner magnetosphere is driven by radial diffusion.

A common understanding
is that the particle number per flux-tube volume tends to be homogenized  (see for example \cite{Has,Schippers}),
hence, a diffusion equation should be written on a magnetic coordinate system.
Transforming back to the ordinary Cartesian coordinates, 
we observe the inward diffusion, because a thinner flux tube near the
magnetic dipole will have a higher density.

Recently, the idea of using the magnetic coordinates was put into the perspective of phase-space foliation (or, degenerate Poisson algebra),
by which a ``macro-scale hierarchy'' was given a mathematical formulation as a \emph{leaf} of Casimir invariants\,\cite{YosFol}.
An adiabatic invariant translates into a Casimir invariant of noncanonical Hamiltonian system that
describes the dynamics of \emph{quasi-particles}, i.e., the representations of
coarse-grained particles ignoring the microscopic action-angle degree of freedom.
The scale of adiabaticity separates macro and micro hierarchies
(reversing the viewpoint, we can interpret a Casimir invariant as some adiabatic invariant, and formulate a 
singular perturbation unfreezing the corresponding topological constraint\,\cite{YosFol2,YosFol3}).
By maximizing the entropy on a symplectic leaf  (Poisson submanifold) of the phase space, 
we can derive a Boltzmann distribution of quasi-particles.
Embedding the leaf into the total phase space on the laboratory frame,
we obtain a clump of magnetized particles that simulates experimental observations
(see \cite{Yos4,Yos,Saitoh}).

The aim of this work is to develop a diffusion model on the foliated phase space.
The stochasticity stems in the macro hierarchy;
we consider a turbulence of collective modes producing a random perturbation $-\nabla\delta\phi$
of macroscopic electric field,
which drives $\boldsymbol{E}\times\boldsymbol{B}$ drifts of the magnetized particles
(we assume that particle collisions can be neglected, so that the magnetic moment $\mu$ is
an adiabatic invariant).
Due to the inhomogeneity of the magnetic field $\boldsymbol{B}=\nabla\psi\times\nabla\theta$
($\psi$ is the magnetic flux function, and $\theta$ is the toroidal angle), this mechanism gives rise to
an inhomogeneous random walk.

There are some delicate problems that need to be carefully addressed in formulating the
Fokker-Planck equation.
They are primarily due to the $\boldsymbol{E}\times\boldsymbol{B}$ drift approximation and the
complexity (non-orthogonality) of the canonical coordinate system.
While the proper coordinate system (bearing an invariant measure) is identified as a symplectic leaf of the
Casimir (adiabatic) invariant, 
the macroscopic Hamiltonian needs a subtle amendment in order to describe
the $\boldsymbol{E}\times\boldsymbol{B}$ drift (see \cite{Cary} as well as footnote 1 of \cite{YosFol}).
Neglecting the kinetic part $mrv_\theta$ of the canonical toroidal momentum $P_\theta=mrv_\theta + q\psi$
($m$ is the mass, $q$ is the charge, and $r$ is the radius),
we approximate $P_\theta \approx q\psi$, and then,
a canonical equation $\dot{\psi}=\partial_\theta \delta\phi$ describes the $\boldsymbol{E}\times\boldsymbol{B}$.
Here, we have used the action variable $\psi$ as a spatial coordinate
(in fact, the $\boldsymbol{E}\times\boldsymbol{B}$ drift velocity, not the acceleration, 
is proportional to the electric force).
To span the configuration space, we introduce a parallel coordinate $\ell$ along each field line (contour of $\psi$).
The cotangent vectors $d\psi$ and $d\ell$ are not orthogonal;
hence, we need an appropriate ``drift energy'' in the Hamiltonian to produce $\dot{\ell}$ associated with the
perpendicular drift velocity.
Instead of following the rather complicated Hamiltonian formalism
to determine the diffusion operator,
we may reverse the process of formulation.
We start by assuming a random $\boldsymbol{E}\times\boldsymbol{B}$ drift velocity $\dot{\boldsymbol{X}}$ 
in the natural coordinates (i.e.,the velocity is a member of the tangent bundle);
we consider a Wiener process, in the perpendicular direction, multiplied by an unknown amplitude.
We transform it to the canonical variables by
invoking Ito's formula of coordinate transformations.
The diffusion coefficient must be made constant on the canonical coordinate system, by which the
amplitude of the Wiener process is determined. 
The derived diffusion equation has an interesting ``drift terms'' describing the geometrical coupling of
the two spatial coordinates $\psi$ and $\ell$.

\section{Invariant Measure} 
We start by reviewing the phase-space foliation due to the constraint of adiabatic invariant (here, the magnetic moment).
We consider a macroscopic system of magnetized particles;
the macro-scale hierarchy is formally defined by separating the microscopic action-angle variables
$\mu$-$\theta_c$ ($\mu=mv_c^2/(2B)$ is the magnetic moment,
$v_c$ is the cyclotron velocity, and $\theta_c$ is the gyration angle) from the
total action of charged particles.
Then, the actual phase space is a sub-manifold (leaf) of the 6-dimensional phase space ($\mu$-space),
which is a symplectic leaf span by the canonical variables
$(\ell, v_\parallel, \psi, \theta)$,
where $\ell$ is the arch-length along magnetic field lines defined by $\partial_{\ell}=\boldsymbol{B}/B$,
$v_{\parallel}$ is the velocity parallel to field lines,
$\psi$ is the toroidal angular momentum (neglecting the mechanical part $m r v_\theta$) and $\theta$
is the toroidal angle. 
Regarding $\mu$ as an attribute of each particle, we consider statistical mechanics on this symplectic leaf.
The natural invariant measure is, after separating the microscopic variables,
\begin{equation}
dV_m= d\ell  dv_{\parallel} d\psi d\theta .
\end{equation}
We consider the Wiener process that tends to maximize the entropy counted by this measure.

Let us see the relation between the laboratory frame and the symplectic leaf of magnetized particles.
The volume element of the former is, in the $(r,z,\theta)$ cylindrical coordinates,
\begin{equation}
dV=rdrdzd\theta dv_{\parallel}dv_{\perp}dv_{\theta} ,
\end{equation}  
where $v_{\perp}$ is the perpendicular velocity, and $v_{\theta}$ is the toroidal velocity. 
Neglecting the toroidal drift velocity
with respect to the cyclotron velocity $v_c$, we 
may approximate $v_\perp = v_c$, and then,
\begin{equation}
dV=rdrdzd\theta dv_{\parallel}v_{c}dv_{c}d\theta_{c}=\frac{B}{m}rdrdzd\theta dv_{\parallel}d\mu d\theta_{c} .
\end{equation}
Using $\vert\nabla \ell\cdot\nabla\psi\times\nabla\theta\vert=B$, we may write
\begin{equation}
dV=\frac{1}{m}d\ell d\psi d\theta dv_{\parallel}d\mu d\theta_{c} .
\end{equation}
On the symplectic leaf, $\psi$ and $\ell$ play the role of spatial coordinates
(the functional forms of $\psi(r,z)$ and $\ell(r,z)$ are given in appendix A).
The spacial volume of the laboratory frame and that of the symplectic leaf are related by
\begin{equation}
d^{3}x=\frac{dV}{dv_{\parallel}dv_{\perp}dv_{\theta}}=\frac{m}{B}\frac{dV}{dv_{\parallel}d\mu d\theta_{c}}
=\frac{2\pi}{B} d\ell d\psi .
\end{equation}
The Jacobian weight $B/2\pi$ is the noted factor that multiplies to the symplectic-leaf density.

\section{Stochastic Motion of Particles}

\subsection{Perpendicular Motion by $\boldsymbol{E}\times\boldsymbol{B}$ Drift}

Here consider electrostatic fluctuations as the causal of transport on the symplectic leaf of magnetized particles.
Upon the $\boldsymbol{E}\times\boldsymbol{B}$ drift approximation (we neglect the
higher-order polarization drift), the velocity in the direction normal to flux surfaces is given by 
\begin{equation}
\boldsymbol{v}_{\perp}=\frac{\delta\boldsymbol{E}\times\boldsymbol{B}}{B^{2}}=\frac{\delta E_{\theta}}{B}\partial_{\perp}=
-\frac{1}{rB}\frac{\partial\delta\phi}{\partial\theta}\partial_{\perp}\label{vper} .
\end{equation}
Here $\partial_{\perp}$ stands for the unit vector in the normal direction:
\begin{equation}
\partial_{\perp}=\frac{\nabla\psi}{\vert\nabla\psi\vert}=\frac{\nabla\psi}{rB} .
\end{equation}
We assume 
\begin{subequations}
\begin{align}
&\langle\delta\phi(t)\rangle=0\label{deltaphi1} ,\\
&\langle\delta\phi(t)\delta\phi(s)\rangle\propto\delta(t-s)\label{deltaphi2} .
\end{align}
\end{subequations}
The standard assumptions (\ref{deltaphi2}) and (\ref{deltaphi2}), respectively, mean 
the average charge neutrality and the de-correlation of the fluctuation at the scale of the present
macroscopic model, 
i.e., $\delta\phi$ is a Gaussian white noise.
The drift velocity $v_{\perp}$ of equation (\ref{vper}) is given by
\begin{equation}
dX_{\perp}=v_{\perp}dt=\frac{D_{\perp}^{1/2}}{rB}dW_{\perp}\label{stocvper} ,
\end{equation}
where $dW_{\perp}=\Gamma_{\perp}(t)dt$ is a Wiener process \cite{Elia} acting in the direction perpendicular to the flux
surfaces, and $D_{\perp}^{1/2}(\boldsymbol{z})$, with $\boldsymbol{z}=(\ell,\psi,v_{\parallel},\mu)$, is 
the magnitude of fluctuations, which will be determined in sec. V.  
We denote random variables by upper-case letters.

\subsection{Parallel Motion}

The equation of motion in the parallel direction is written as

\begin{equation}
m\frac{dv_{\parallel}}{dt}=-\mu\frac{\partial B}{\partial \ell}-q\frac{\partial\delta\phi}{\partial \ell}
-m\gamma v_{\parallel} .
\end{equation}

\noindent The first term on the right hand side is the mirror force. 
The third term is a friction force ($\gamma$ is the friction coefficient).  
The parallel electric field is usually much smaller than the perpendicular one.
We assume

\begin{equation}
-\frac{q}{m}\frac{\partial\delta\phi}{\partial\ell}=D_{\parallel}^{1/2}\frac{dW_{\parallel}}{dt} ,
\end{equation}

\noindent The amplitude $D_{\parallel}^{1/2}$ is a constant since $v_{\parallel}$ is a canonical
variable on which the diffusion coefficient has to be Fickian. 
 
The stochastic equation of motion in the parallel direction reads

\begin{equation}
dV_{\parallel}=-\left(\frac{\mu}{m}\frac{\partial B}{\partial \ell}+\gamma v_{\parallel}
\right)dt+D_{\parallel}^{1/2}dW_{\parallel} .\label{vlSDE}
\end{equation}

The stochastic evolution of the coordinate $\ell$ (which will be denoted by $L$) is not simply the time integral of
$v_{\parallel}$, because particles are moving on the curvilinear coordinates: due to $\nabla \ell\cdot\partial_{\perp}\not=0$
a velocity in $\partial_{\perp}$ also causes a change in $\ell$. Including this effect, we obtain (see appendix D for the derivation)

\begin{dmath}
dL=\left\{v_{\parallel}+\left(\frac{1}{2}-\alpha\right)D_{\perp}\left[\left(\mathfrak{q}\frac{\partial}{\partial \ell}+\frac{\partial}{\partial\psi}\right)\mathfrak{q}+\mathfrak{q}\left(\mathfrak{q}\frac{\partial}{\partial \ell}+\frac{\partial}{\partial\psi}\right)\ln(rB)\right]\right\}dt+\mathfrak{q}D_{\perp}^{1/2}dW_{\perp} ,\label{l}
\end{dmath}

\noindent where the geometrical factor $\mathfrak{q}$ is defined by

\begin{equation}
\mathfrak{q}=\frac{\nabla \ell\cdot\nabla\psi}{(rB)^{2}} .
\end{equation}

The coefficient $\alpha$ will be explained in the next section.

\section{Stochastic Integrals}

In the previous section, we have derived a system of stochastic differential equations (SDEs) (\ref{stocvper}), (\ref{vlSDE}), and (\ref{l}). To obtain the corresponding Fokker-Planck equation (FPE) we need a rigorous definition of the stochastic integral that determines the solution to our equations of motion. Indeed, 
While the deterministic terms (multiplied by $dt$)
can be integrated by the usual Riemann integral, the random terms (multiplied by $dW$)
require careful treatment.  
One may integrate a stochastic term differently depending on the choice of $t_{\alpha}\in[t_{i-1},t_{i}]$ 
 ($t_{i-1}$ and $t_{i}$ are neighbouring times) \cite{Gar,Ris,Evans}.
Let us do a practice with the perpendicular equation (\ref{stocvper}).
If we put $t_{\alpha}=t_{i-1}+\alpha\Delta t_{i}$ with $\alpha\in[0,1]$,
approximate the integral 
with a finite sum, and then take the limit 
for $\Delta t_{i}=\frac{t-t_{0}}{n}\rightarrow 0$ as in the Riemann integral, we have

\begin{equation}
\begin{split}
\int_{t_{0}}^{t}dX_{\perp}=
ms\text{-}\lim_{n \to \infty}\sum_{i=1}^{n}&\frac{D_{\perp}^{1/2}}{rB}\left(\boldsymbol{X}(t_{\alpha})\right)\\
&[W_{\perp}(t_{i})-W_{\perp}(t_{i-1})] .\label{SI1}
\end{split}
\end{equation}
 
\noindent Here "$ms$" stands for mean-square limit 
and $\boldsymbol{X}(s)$ denotes the position in space at time $s$. Note that $\boldsymbol{X}$ is a random process. Because of the nowhere differentiability \cite{Evans} of the Wiener process, the value of the integral (\ref{SI1}) depends on $\alpha$. This problem is known as the \textit{Ito-Stratonovich dilemma} (the reader is referred to \cite{Wong,Kampen,Kampen2,Kuroiwa,Moon,Sancho,Chechkin,Ran,Yuan,Farago,Volpe,Mannella,Lau}). In appendix C, we give a theorem which establishes the relationship between the Ito integral $\alpha=0$ \cite{Gar,Ris,Evans,Ito1,Ito3,Ito4} and that calculated with a different $\alpha$. As an alternative to Ito's definition, the Stratonovich integral assumes $\alpha=1/2$ \cite{Gar,Ris,Evans}.

In the case of homogeneous random walks, all the definitions of the stochastic integral are equivalent (see appendix C). 
However, in an inhomogeneous random walk, FPE depends on $\alpha$. 

\begin{figure}[h]
\centering
\includegraphics[scale=0.4]{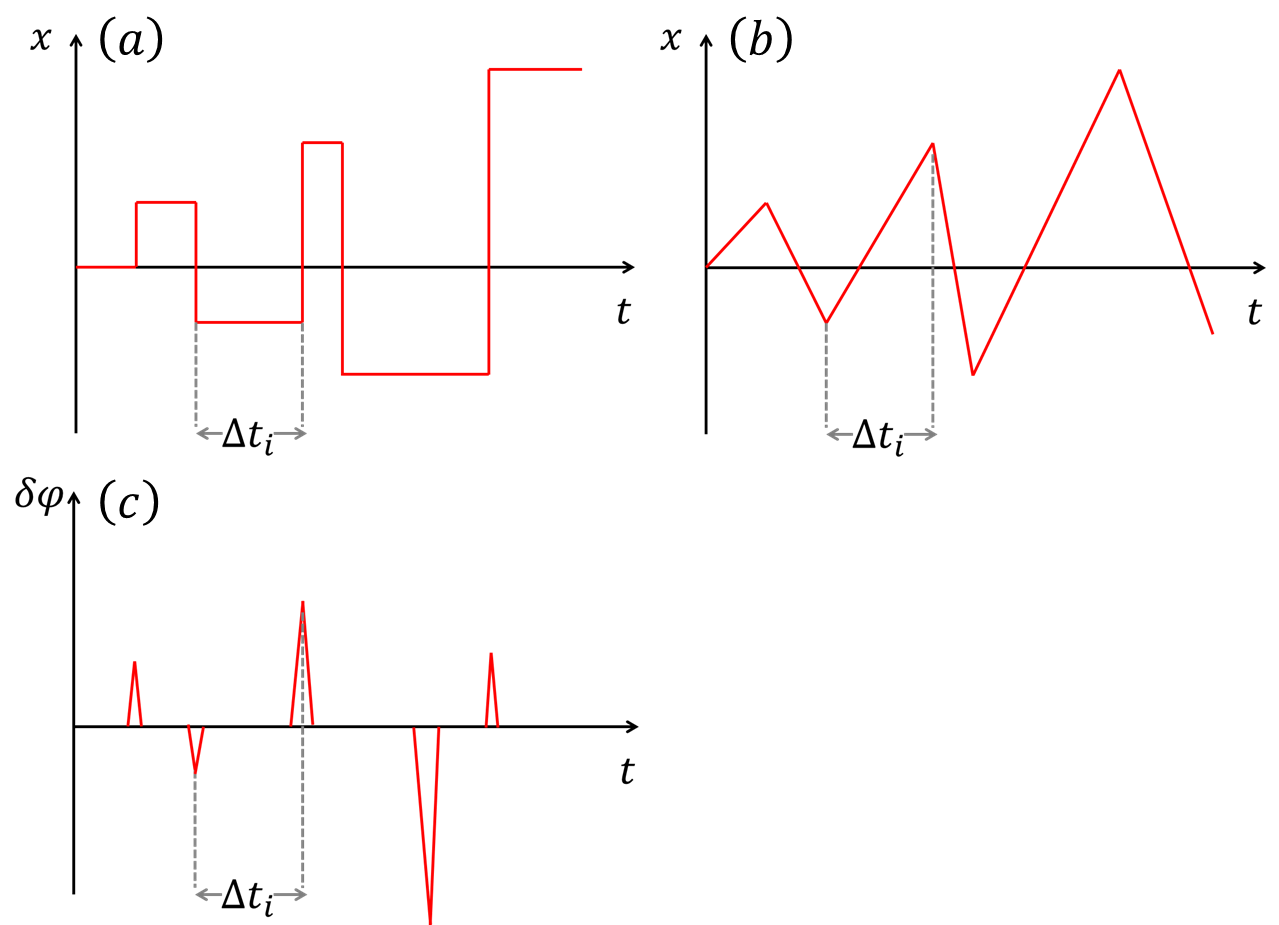}
\caption{\footnotesize (a): schematic representation of the process $\mathcal{P}_{\lambda}\otimes N$. In the limit
$\lambda\rightarrow\infty$ it converges to the Wiener process equipped with Ito's integral. (b): schematic representation of the process $\mathcal{P}_{\lambda}\otimes N'$. In the limit $\lambda\rightarrow\infty$ it converges to the Wiener process equipped with Stratonovich integral $\alpha=1/2$. (c): schematic representation of the turbulent electric potential $\delta\phi$.}
\label{figW}
\end{figure}

We have to choose an appropriate $\alpha$ by examining the physical information contained in $\alpha$.
The Ito integral with $\alpha=0$ assumes
that two neighboring values of the Wiener process are connected by a step function.
More precisely, if we combine a Poisson process $\mathcal{P}_{\lambda}$, which samples
the times at which fluctuations occur (with rate $\lambda$), and a Gaussian process $N(0,t)$, which samples
the amplitude of the fluctuations, we have a process that converges to the Wiener process $W$
equipped with Ito's definition of the stochastic integral in the limit $\lambda\rightarrow\infty$.  
In fact, the integral of a function $f$ over the process $\mathcal{P}_{\lambda}\otimes N$ is
(see figure \ref{figW}a)

\begin{dmath}
I=ms\text{-}\lim_{n\rightarrow\infty}\sum_{i=1}^{n}f(t_{i-1})(W(t_{i})-W(t_{i-1})) ,
\end{dmath}

\noindent and the probability distribution $P_{\lambda}(x,t)$ of having an amplitude $x$ of the
fluctuations at time $t$ converges to a normal distribution $N(0,t)$ with variance $t$:

\begin{dmath}
P_{\lambda}(x,t)=\frac{1}{\sqrt{2\pi t}}e^{-x^{2}/2t}(1-e^{-\lambda t})+\delta(x)e^{-\lambda t}\xrightarrow{\lambda\rightarrow \infty}\frac{1}{\sqrt{(2\pi t)}}e^{-x^{2}/2t} .
\end{dmath}

Consider now a new process $\mathcal{P}_{\lambda}\otimes N'$ where this time $N'(0,t)$ is a normal distribution 
that samples the angular coefficient. The integral of a function $f$ over this new process is

\begin{equation}
I'=ms\text{-}\lim_{n\rightarrow\infty}\sum_{i=1}^{n}f\left(t_{i-1}+\frac{\Delta t_{i}}{2}\right)(W(t_{i})-W(t_{i-1})) ,
\end{equation}

\noindent and the probability distribution $P'_{\lambda}(x,t)$ of having an amplitude $x$ of the
fluctuations at time $t$ converges to a normal distribution $N(0,t)$ with variance $t$ (the proof is given
by representing the process as $x(t)=x_{i-1}+\frac{t-t_{i-1}}{t_{i}-t_{i-1}}(x_{i}-x_{i-1})$ and verifying that its variance goes to $t$ in the limit $\lambda\rightarrow\infty$).
A schematic representation of this new process is given in figure \ref{figW}b.

In bounce motion, a particle experiences
random kicks by the turbulent potential $\delta\phi$ (see figure \ref{figW}c),
which are the temporal derivatives of Ito's $W$ shown in figure \ref{figW}a.
Indeed, $\delta\phi\propto\Gamma(t)=dW/dt$  (see (\ref{deltaphi1}) and (\ref{deltaphi2})).
Hence, we choose $\alpha=0$, i.e., Ito's definition of the stochastic integral.

\section{Coordinate Transformation to the Proper Frame}

For a diffusion process to homogenize the probability distribution with respect to the invariant measure,
the diffusion coefficient (the magnitude of the white noise) must be constant on the symplectic leaf (proper frame);
in (\ref{stocvper}), we have to change the coordinate $X_\perp$ to $\Psi$:

\begin{equation}
d\Psi=Adt+D_{\psi}^{1/2}dW_{\perp}\label{Psi0} ,
\end{equation}

\noindent with $D_{\psi}^{1/2}$ a constant diffusion coefficient and $A$ a drift current.

To evaluate $A$ we apply Ito's lemma of changing variables (from $X_{\perp}$ to $\Psi$) of stochastic calculus \cite{Gar,Ris,Evans}
(since the rule for an arbitrary $\alpha$ is not found in the literature, we derive it in appendix B).
By equations (\ref{stocvper}), (\ref{Psi0}), and (\ref{GIL2}), we obtain

\begin{subequations}
\begin{align}
&A=\left(\frac{1}{2}-\alpha\right)\frac{D_{\perp}}{(rB)^{2}}\frac{\partial^{2}\psi}{\partial x_{\perp}^{2}} ,\\
&D_{\psi}^{1/2}=\frac{D_{\perp}^{1/2}}{rB}\left(\frac{\partial \psi}{\partial x_{\perp}}\right)\label{xPtoY} .
\end{align}
\end{subequations}

\noindent By

\begin{equation}
\frac{\partial\psi}{\partial x_{\perp}}=\nabla\psi\cdot\partial_{\perp}=\frac{\nabla\psi\cdot\nabla\psi}{rB}=rB ,
\end{equation}

\noindent we obtain $D_{\perp}^{1/2}=D_{\psi}^{1/2}$ = constant, and

\begin{dmath}
A=\left(\frac{1}{2}-\alpha\right)\frac{D_{\perp}}{(rB)^{2}}\frac{\partial }{\partial x_{\perp}}rB=
\left(\frac{1}{2}-\alpha\right)\frac{D_{\perp}}{(rB)^{2}}\left(\frac{\partial \ell}{\partial x_{\perp}}\frac{\partial}{\partial \ell}
+\frac{\partial \psi}{\partial x_{\perp}}\frac{\partial}{\partial\psi}\right)rB=
\left(\frac{1}{2}-\alpha\right)\frac{D_{\perp}}{(rB)^{2}}\left(\frac{\partial \ell}{\partial x_{\perp}}\frac{\partial}{\partial \ell}
+rB\frac{\partial}{\partial\psi}\right)rB .\label{A1}
\end{dmath}

\noindent We have 

\begin{equation}
\frac{\partial \ell}{\partial x_{\perp}}=\nabla \ell\cdot\partial_{\perp}=\frac{\nabla \ell\cdot\nabla\psi}{rB} .\label{lxP}
\end{equation}

\noindent Combining (\ref{A1}) and (\ref{lxP}), we obtain

\begin{equation}
A=D_{\perp}\left(\frac{1}{2}-\alpha\right)\left(\frac{\nabla \ell\cdot\nabla\psi}{(rB)^{2}}\frac{\partial}{\partial \ell}
+\frac{\partial}{\partial\psi}\right)\ln\left(rB\right) .\label{PsiCurr}	
\end{equation}

\noindent In summary, 

\begin{dmath}
d\Psi=D_{\perp}\left(\frac{1}{2}-\alpha\right)\left(\mathfrak{q}\frac{\partial}{\partial \ell}+\frac{\partial}{\partial\psi}\right)\ln\left(rB\right)dt\\+D_{\perp}^{1/2}dW_{\perp} .\label{Psi} 
\end{dmath}

\section{Fokker-Planck Equation}
Given a system of SDEs such as

\begin{equation}
dX_{i}=F_{i}dt+G_{ij}dW_{j} ,\label{MultiSDE_}
\end{equation}\\

\noindent and the stochastic integral is defined with an appropriate $\alpha$
(we assume that $\alpha$ is common for all Wiener processes $W_{j}$),
the Fokker-Planck equation governing the probability density $f$ is \cite{Gar,Ris}

\begin{equation}
\frac{\partial f}{\partial t}=\frac{\partial}{\partial x_{i}}\left(-F_{i}+\frac{1}{2}\frac{\partial}{\partial x_{j}}G_{ik}G_{jk}-\alpha\frac{\partial G_{ik}}{\partial x_{j}}G_{jk}\right)f .\label{FPE}
\end{equation}\\

Using (\ref{vlSDE}), (\ref{l}) and (\ref{Psi}) in (\ref{FPE}), we obtain

\begin{dmath}
\frac{\partial P}{\partial t}=-v_{\parallel}\frac{\partial P}{\partial \ell}+\frac{\partial}{\partial v_{\parallel}}
\left[\left(\frac{\langle\mu\rangle_{\boldsymbol{v}}}{m}\frac{\partial B}{\partial \ell}+\gamma v_{\parallel}\right)P\right]
-\left(\frac{1}{2}-\alpha\right)D_{\perp}\frac{\partial}{\partial\psi}\left\{\left[\left(\mathfrak{q}\frac{\partial}{\partial \ell}+\frac{\partial}{\partial\psi}\right)\ln{(rB)}\right]P\right\}-\left(\frac{1}{2}-\alpha\right)D_{\perp}\frac{\partial}{\partial\ell}\left\{\left[\mathfrak{q}\left(\mathfrak{q}\frac{\partial}{\partial \ell}+\frac{\partial}{\partial\psi}\right)\ln{(rB)}\right]P\right\}
-\frac{1}{2}D_{\perp}\frac{\partial}{\partial \ell}\left\{\left[\left(\mathfrak{q}\frac{\partial}{\partial \ell}+\frac{\partial}{\partial\psi}\right)\mathfrak{q}\right]P\right\}+\frac{1}{2}D_{\perp}\frac{\partial^{2}}{\partial \ell^{2}}\left(\mathfrak{q}^{2}P\right)
+D_{\perp}\frac{\partial^{2}}{\partial \ell\partial\psi}\left(\mathfrak{q}P\right)+\frac{1}{2}D_{\perp}\frac{\partial^{2}P}{\partial\psi^{2}}+\frac{1}{2}D_{\parallel}\frac{\partial^{2}P}{\partial v_{\parallel}^{2}} .\label{FPEIWComplete}
\end{dmath}

\noindent Here $P$ is the probability distribution on the proper frame $(\ell,\psi,v_{\parallel})$,
and $\langle\mu \rangle_{\boldsymbol{v}}$ is the average of $\mu$ over the velocity space. 
The laboratory-frame spatial density $\rho$ is given by 

\begin{equation}
\rho=\frac{d\ell d\psi d\theta}{rdrdzd\theta}\int{Pdv_{\parallel}}=\vert\nabla \ell\cdot\nabla\psi\times\nabla\theta\vert u=Bu ,
\end{equation}

\noindent where $u=\int{Pdv_{\parallel}}$ is the spatial density on the proper frame. 
It is now evident in (\ref{FPEIWComplete}) that the diffusion on the coordinate $\psi$ 
is Fickian with a constant diffusion coefficient, transporting particles to diminish 
the gradient with respect to $\psi$.

\section{Numerical Simulation}

We put the formulated model to the test by numerical simulation.
For the purpose of proving the principle, we consider a simple geometry generated by a point dipole
magnetic field.

\subsection{Model}

To simplify the calculations we make the following approximation:

\begin{equation}
dL\simeq v_{\parallel}dt+\mathfrak{q}D_{\perp}^{1/2}dW_{\perp} ,
\end{equation}

\noindent i.e., the transport current appearing in (\ref{l}) is considered to be
much smaller than the convective drift $v_{\parallel}$.
Then, FPE becomes

\begin{dmath}
\frac{\partial P}{\partial t}=-v_{\parallel}\frac{\partial P}{\partial \ell}+\frac{\partial}{\partial v_{\parallel}}
\left[\left(\frac{\langle\mu\rangle_{\boldsymbol{v}}}{m}\frac{\partial B}{\partial \ell}+\gamma v_{\parallel}\right)P\right]
-\frac{1}{2}D_{\perp}\frac{\partial}{\partial\psi}\left\{\left[\left(\mathfrak{q}\frac{\partial}{\partial \ell}+\frac{\partial}{\partial\psi}\right)\ln{(rB)}\right]P\right\}+\frac{1}{2}D_{\perp}\frac{\partial^{2}}{\partial \ell^{2}}\left(\mathfrak{q}^{2}P\right)
+D_{\perp}\frac{\partial^{2}}{\partial \ell\partial\psi}\left(\mathfrak{q}P\right)+\frac{1}{2}D_{\perp}\frac{\partial^{2}P}{\partial\psi^{2}}+\frac{1}{2}D_{\parallel}\frac{\partial^{2}P}{\partial v_{\parallel}^{2}}\label{FPEIWIto} ,
\end{dmath}

\noindent where we put $\alpha=0$ as concluded in section IV.

To solve (\ref{FPEIWIto}), we need to specify $\langle\mu\rangle_{\boldsymbol{v}}$.
Here we assume a local Boltzmann distribution as derived in \cite{Yos}.
Since $\mu$ is a robust adiabatic invariant, we may consider a grand canonical ensemble 
determined by the energy and magnetization (total magnetic moment).
Maximizing the entropy, we obtain

\begin{equation}
f^{eq}=\Theta(\ell,\psi,v_{\parallel})e^{-\beta(B+\zeta)\mu} ,\label{feq}
\end{equation}

\noindent where $\zeta$ is the Lagrangian multiplier (chemical potential) with respect to the magnetization,
$\beta$ the inverse temperature, and $\Theta$ a function independent of $\mu$. Then,

\begin{equation}
\langle\mu\rangle_{\boldsymbol{v}}^{eq}=\frac{\int{\mu f^{eq}d\mu}}{\int{f^{eq}d\mu}}=\frac{1}{\beta(B+\zeta)} .
\end{equation}

\subsection{Results of Numerical Simulations}
Here we study the free decay of the density starting from an initial
Maxwell-Boltzmann distribution in the laboratory frame:

\begin{equation}
P_{lab}(t=0)=BP(t=0)=\sqrt{\frac{m\beta}{2\pi}}\rho_{0}e^{-\beta\frac{m}{2}v_{\parallel}^{2}} .
\end{equation}

\noindent The corresponding spatial density is, of course, constant ($=\rho_{0}$) in the calculation domain,
which is surrounded by a magnetic surface (a level set of $\psi$) as well as a magneto-isobaric surface 
near the origin $(r,z)=(0,0)$. 
We assume a homogeneous Dirichlet boundary condition.

We assume that the parallel electric field is much (three orders of magnitude) smaller than
the perpendicular electric field. In particular, the characteristic normal and parallel fluctuating potentials
are $\delta\phi_{\perp}=500V$ and $\delta\phi_{\parallel}=1V$ acting on typical time scales
of $\tau_{\perp}=10\mu s$ and $\tau_{\parallel}=10 ns$ so that the non-normalized\footnote{In the numerical
simulation, the FPE is normalized through the characteristic quantities $B_{0}=0.1 T$, $\rho_{0}=10^{17}m^{-3}$, $L_{0}=0.25 m$, $\beta^{-1}=1keV$. These values are representative of the RT-1 device \cite{Yos}.} diffusion parameters
are $D_{\perp}=\delta\phi_{\perp}^{2}/\tau_{\perp}=2.5\, 10^{10}\, V^{2}s^{-1}$ and $D_{\parallel}=(q\delta\phi_{\parallel})^{2}\tau_{\parallel}/(m\Delta\ell)^{2}=3.1\, 10^{24} m^{2}s^{-3}$, where we assumed a typical fluctuation length scale 
$\Delta\ell=10\mu m$.

\begin{figure}[h]
\centering
\includegraphics[scale=0.4]{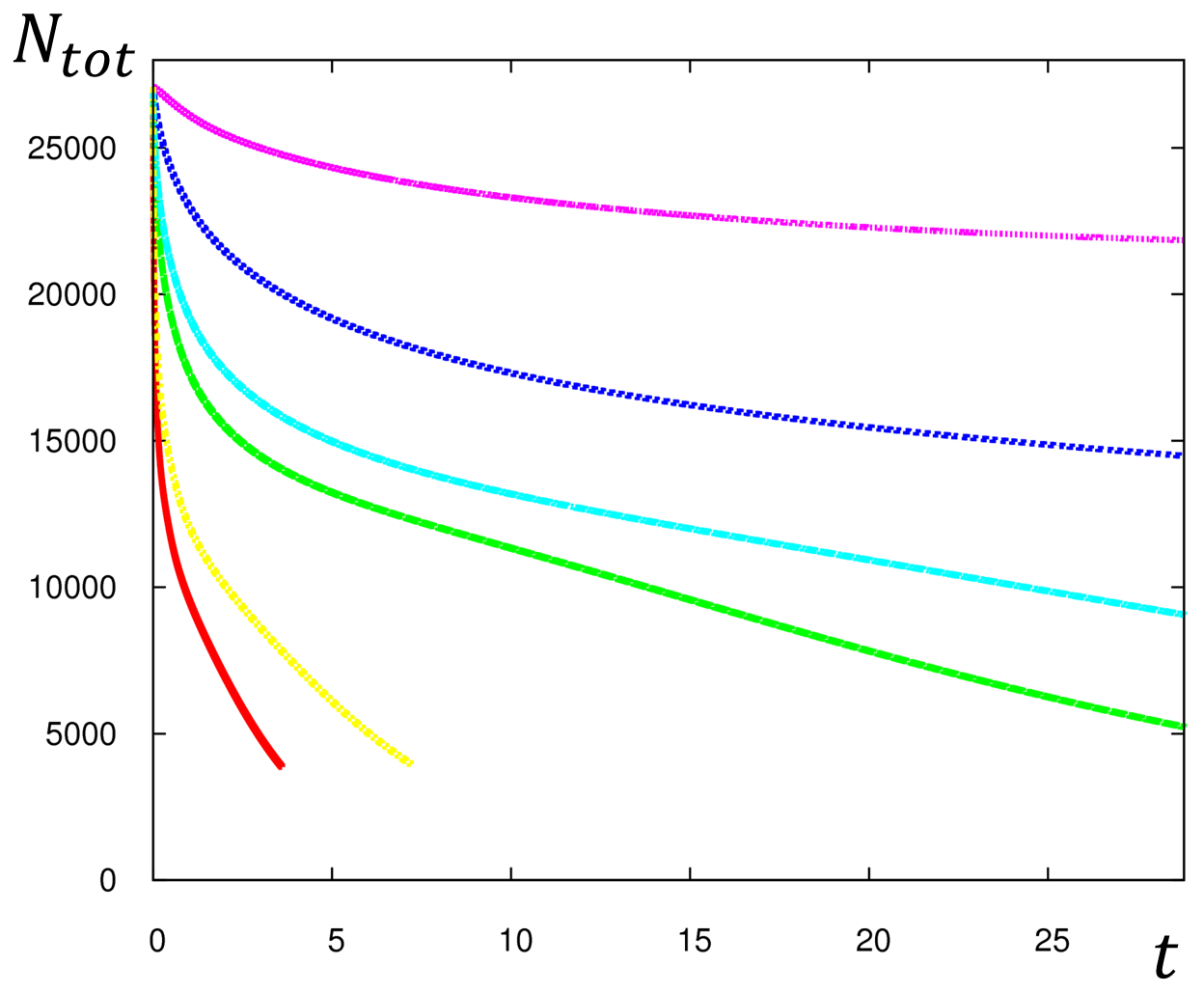}
\caption{\footnotesize Time evolution of the total particle number $N_{tot}$ in arbitrary units. Time $t$
is calculated as number of numerical steps times the used time step. Different colors correspond to different values of
the normal diffusion parameter $D_{\perp}$: purple $10^{-2}D_{\perp}$, blue $10^{-1}D_{\perp}$, sky blue $0.5D_{\perp}$, green $D_{\perp}$, yellow $5D_{\perp}$, and red $10D_{\perp}$.}
\label{fig4}
\end{figure}

Figure \ref{fig4} shows the change of the total number of particles:

\begin{equation}
N_{tot}=\int_{\partial V}{Pd\ell d\psi dv_{\parallel}} .
\end{equation}

\noindent Because of the absorbing boundaries, the number of particles contained in the
domain decreases. 

\begin{figure}[h!]
\centering
\includegraphics[scale=0.5]{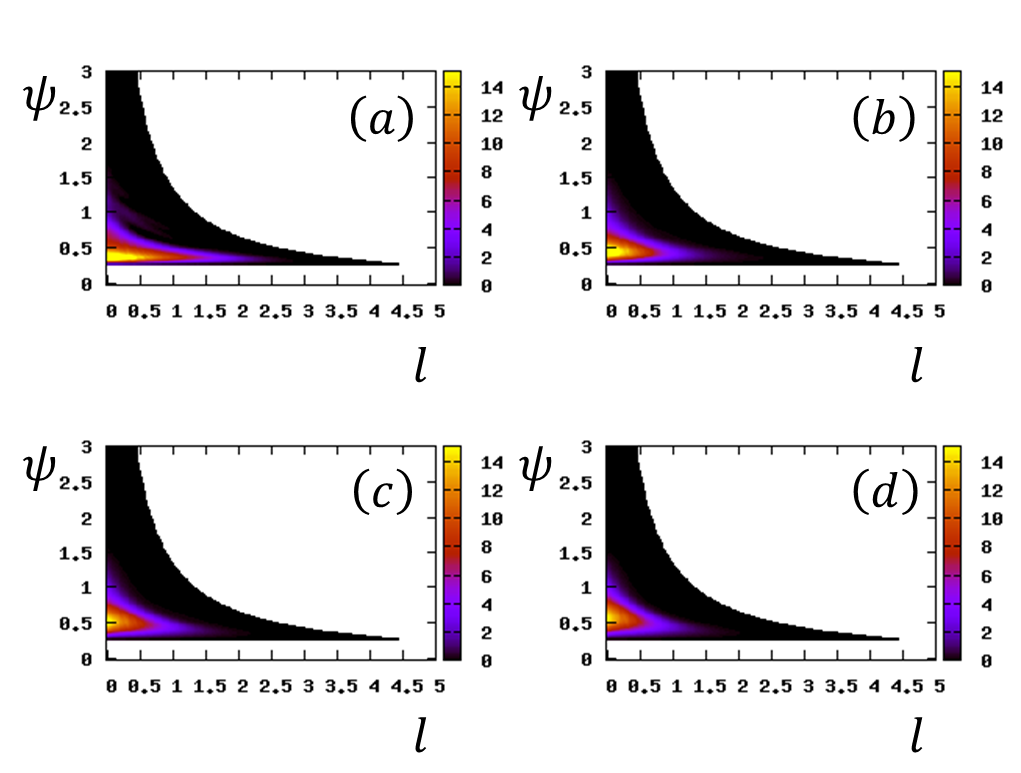}
\caption{\footnotesize Time evolution of the proper-frame density $u$. (a): $t=0.24$. (b): $t=0.96$. (c): $t=1.68$. (d): $t=2.4$.}
\label{fig5}
\end{figure}

\begin{figure}
\centering
\includegraphics[scale=0.45]{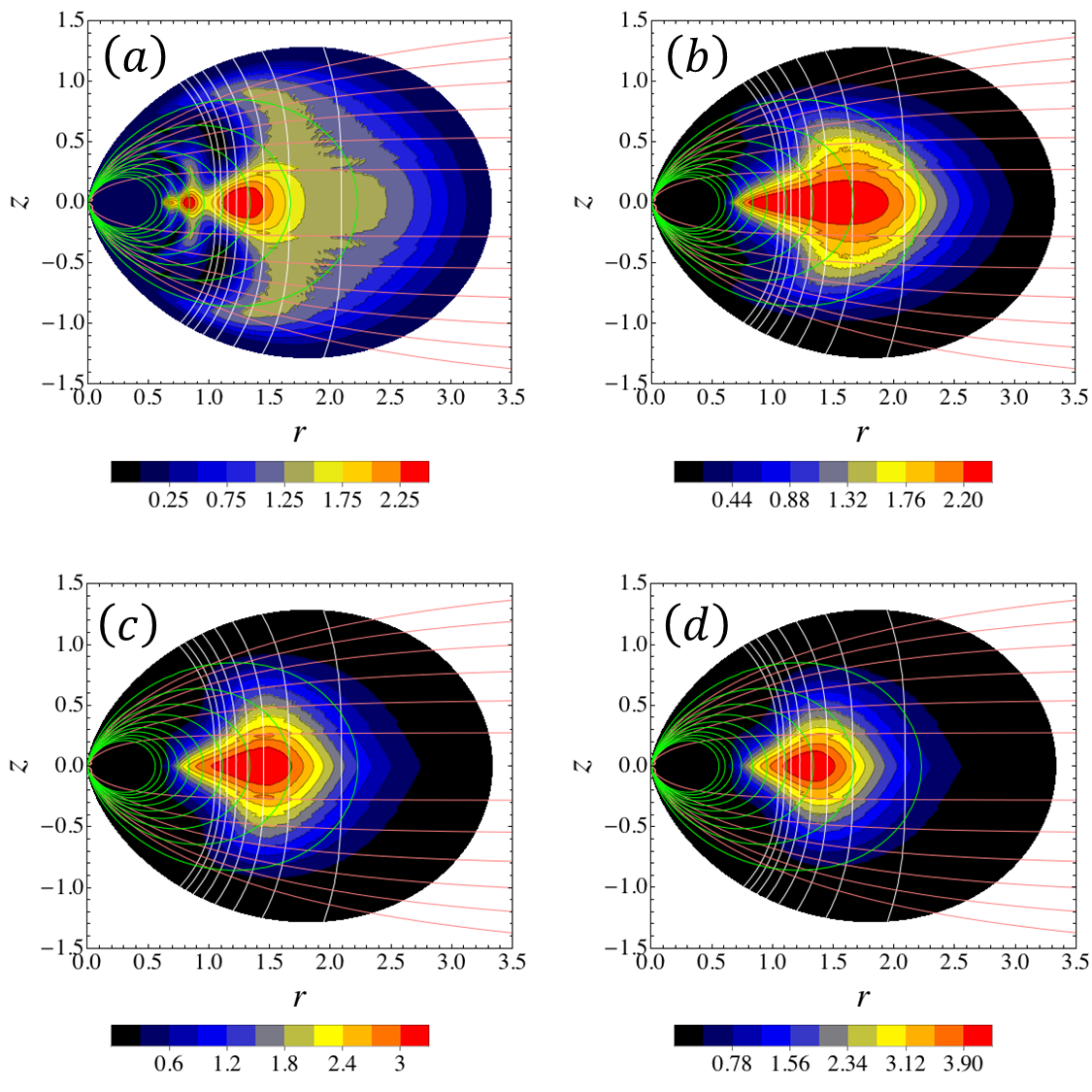}
\caption{\footnotesize Time evolution of the laboratory frame density $\rho$. (a): $t=0.24$. (b): $t=0.96$. (c): $t=1.68$. (d): $t=2.4$.}
\label{fig6}
\end{figure}

Figure \ref{fig5} shows the evolution of the particle density $u$ on the proper frame.
Particles are gradually squeezed into the equatorial region;
this is due to the mirror effect.
At the same time, the profile of $u$ broadens in the direction of $\psi$.
Transforming back to the laboratory frame, the corresponding evolution of $\rho$ is viewed as inward diffusion
(creation of a density gradient);
see figure \ref{fig6}.
In the figures, green curves are magnetic field lines (contours of $\psi$), pink curves are the parallel
($\ell$) coordinates, and white curves are the contours of the magnetic field strength.

\begin{figure}
\centering
\includegraphics[scale=0.45]{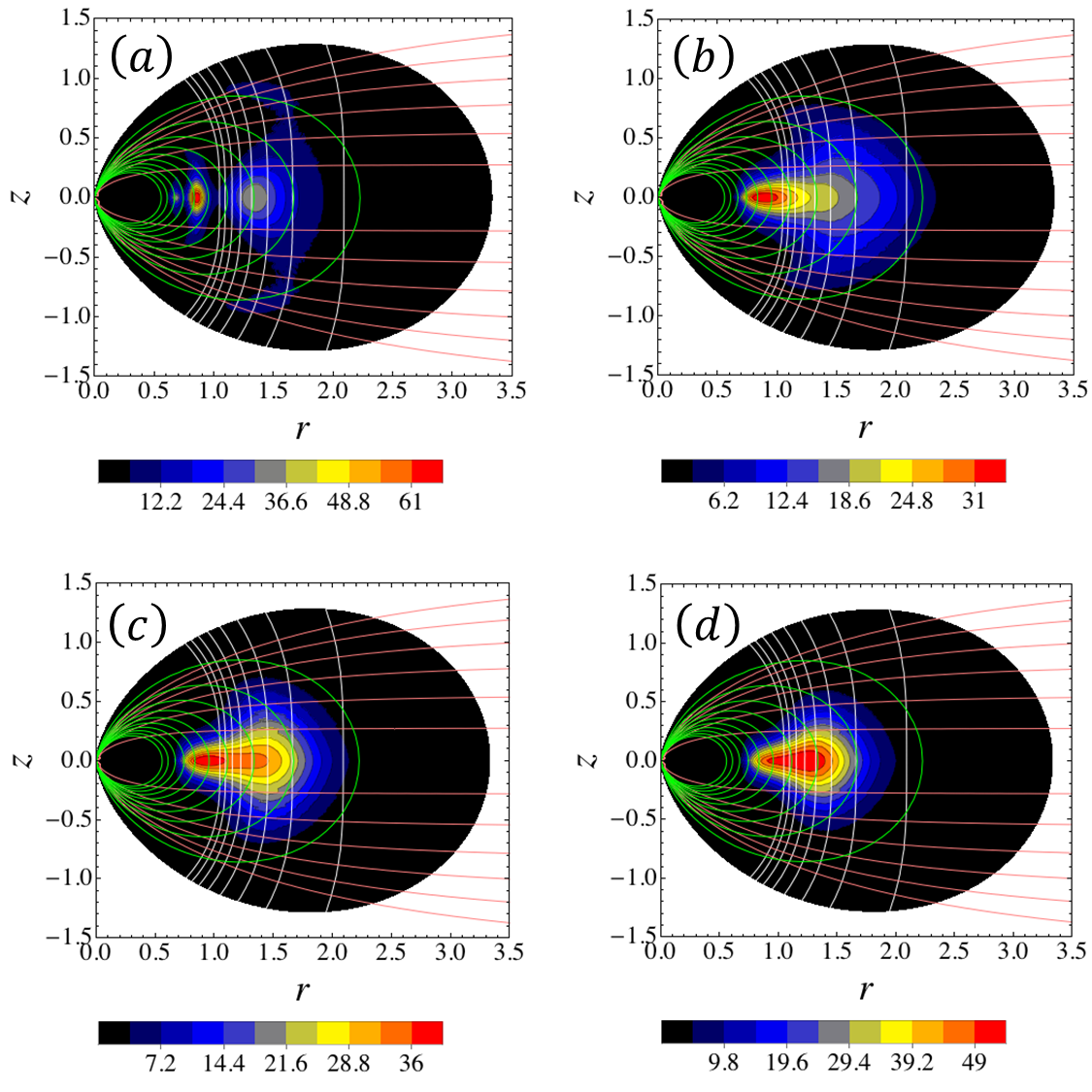}
\caption{\footnotesize Time evolution of the pressure $\mathcal{P}_{\parallel}$ in the laboratory frame.
(a): $t=0.24$. (b): $t=0.96$. (c): $t=1.68$. (d): $t=2.4$.}
\label{fig7}
\end{figure}

Figure \ref{fig7} shows the evolution of the parallel pressure:

\begin{equation}
\mathcal{P}_{\parallel}=\rho \left\langle \left(v_{\parallel}-\langle v_{\parallel}\rangle\right)^{2}\right\rangle .
\end{equation}

\noindent The ``confinement'' in the parallel direction is due the the magnetic mirror effect.

\begin{figure}
\centering
\includegraphics[scale=0.45]{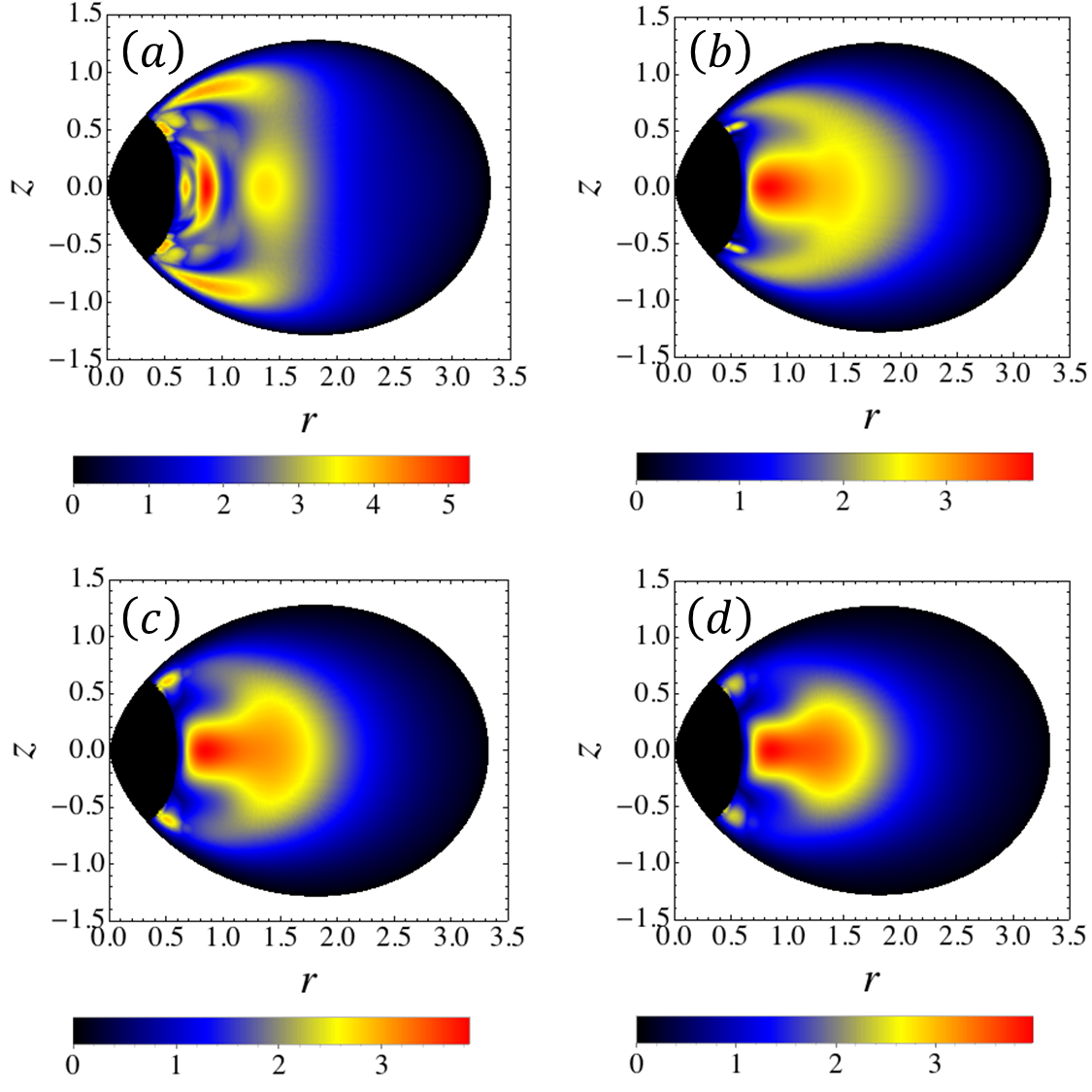}
\caption{\footnotesize Time evolution of the root mean square velocity $\langle v_{\parallel}^{2}\rangle$ in the laboratory frame. (a): $t=0.24$. (b): $t=0.96$. (c): $t=1.68$. (d): $t=2.4$.}
\label{fig9}
\end{figure}

In figure \ref{fig9}, we plot the root mean square parallel velocity: 

\begin{equation}
\sqrt{\langle v_{\parallel}^{2}\rangle}=\sqrt{\frac{\int{v_{\parallel}^{2}Pdv_{\parallel}}}{\int{Pdv_{\parallel}}}}=\sqrt{\frac{1}{u}\int{v_{\parallel}^{2}Pdv_{\parallel}}} ,
\end{equation}

\noindent which is a measure of the parallel temperature $T_{\parallel}=\frac{m}{2}\langle v_{\parallel}^{2}\rangle$.
We observe higher temperatures near the equator.

\begin{figure}[h!]
\centering
\includegraphics[scale=0.45]{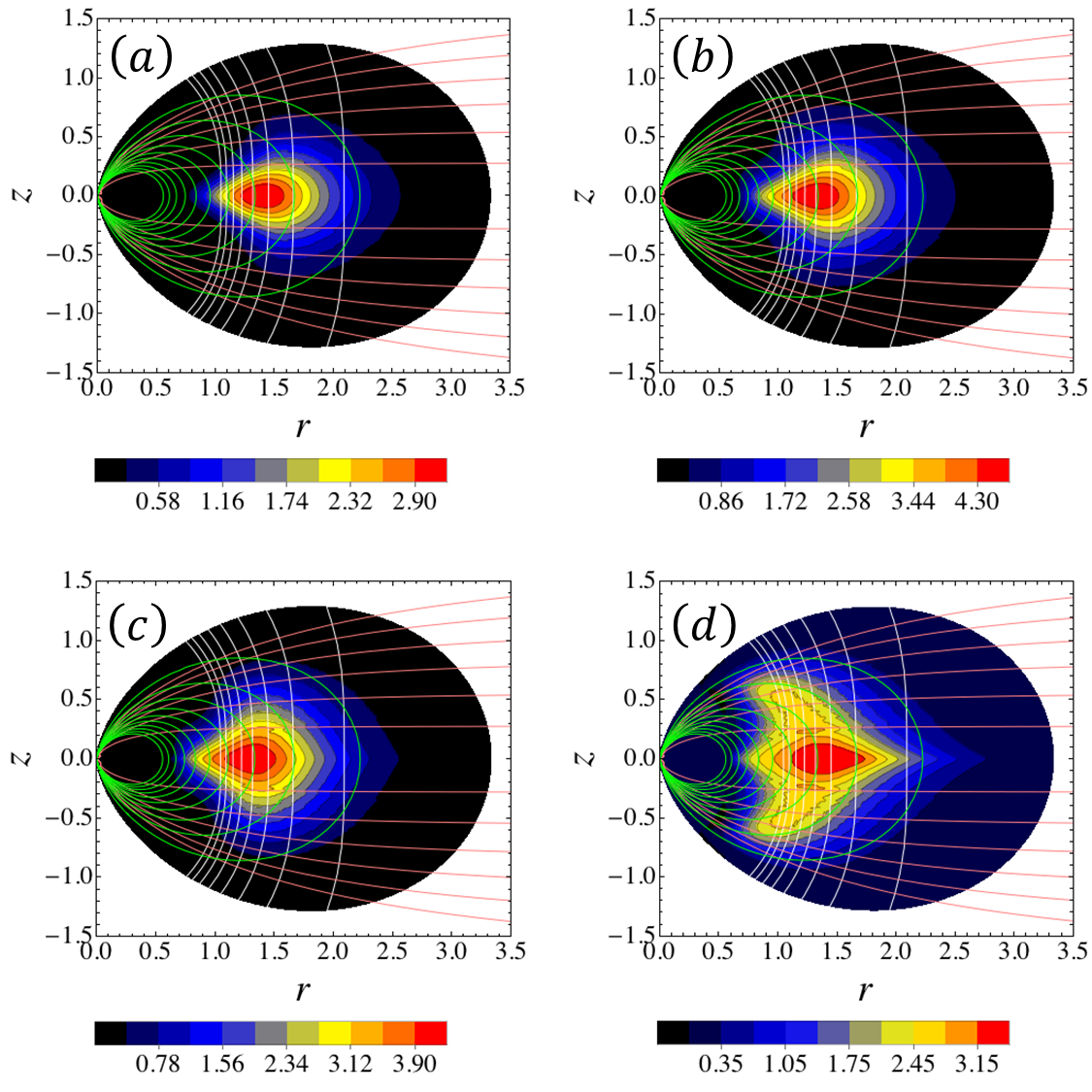}
\caption{\footnotesize Temperature dependence of the density $\rho$ in the laboratory frame at $t=2.4$. 
(a): $10^{-2}\beta^{-1}$. (b): $10^{-1}\beta^{-1}$. (c): $\beta^{-1}$. (d): $10\beta^{-1}$}
\label{fig11}
\end{figure}

A higher initial temperature $\beta^{-1}$ enables particles to have larger bounce orbits.
Consequently, the density may have a wider distribution along field lines. 
In figure \ref{fig11}, we compare the density distributions (after the same passed) for different initial temperatures.

In figure \ref{fig4}, we compare the evolutions with different diffusion coefficients $D_{\perp}$,
which, in (\ref{FPEIWIto}), controls the relative strength of the diffusion effect 
with respect to the parallel mirror effect.
Usually, a stronger diffusion yields a flatter distribution of particles.
Here, the opposite is the case in the laboratory frame; a larger $D_{\perp}$ decreases the
gradient $\partial P/ \partial\psi$ more rapidly, resulting in a faster and stronger steepening of the density $\rho$
in the laboratory frame.
In figure \ref{figRM}, we plot the peak density $\rho_{M}$ (normalized by $\rho_0$)
and the radial position of the peak as functions of time.
When $D_{\perp}$ is increased, the instantaneous maximum density, appearing at smaller radius, experiences a larger value,
while the total particle number $N_{tot}$ diminishes faster (see figure \ref{fig4}). 
Similar behavior is observed in the pressure peak; figure \ref{figPM}.

\begin{figure}[h!]
\centering
\includegraphics[scale=0.46]{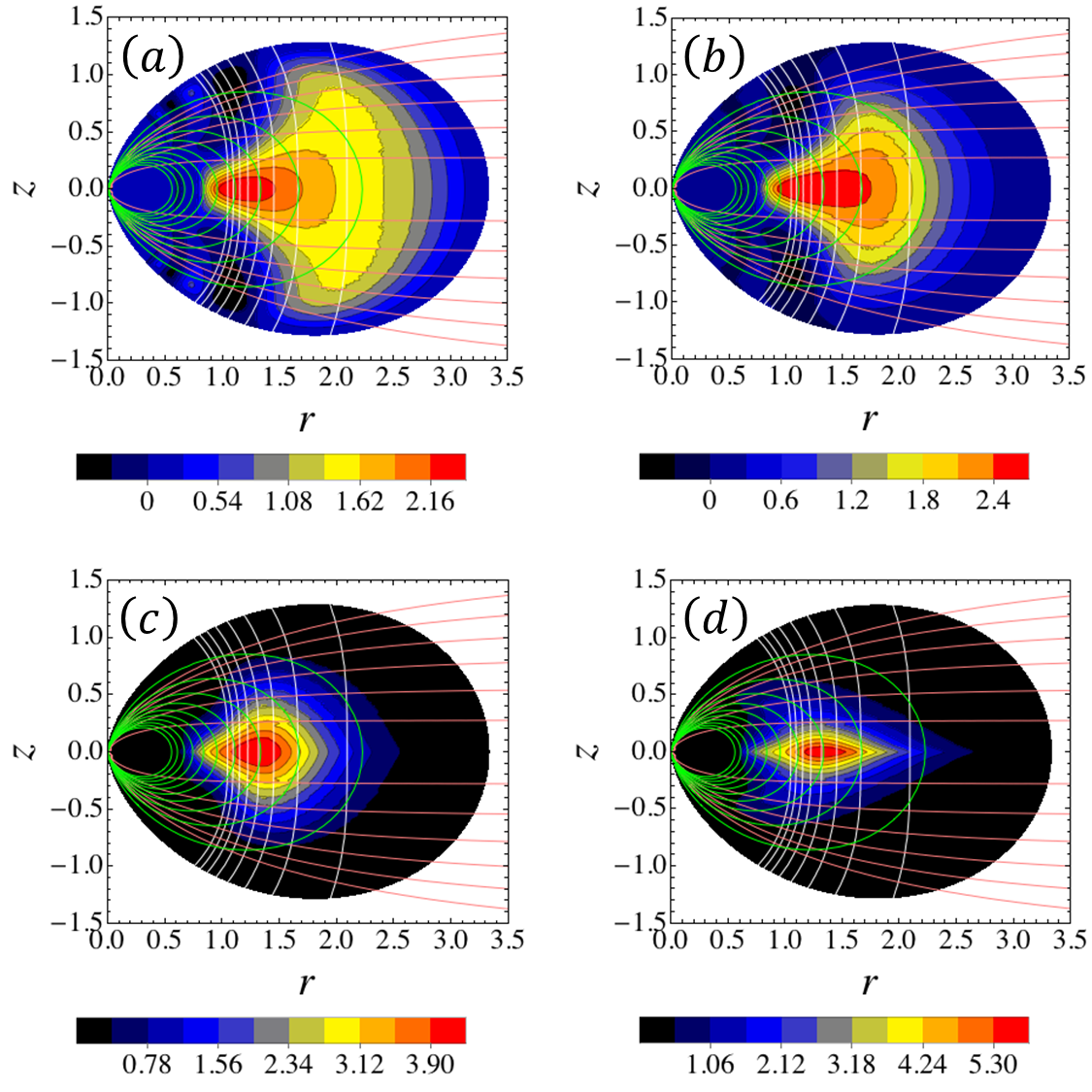}
\caption{\footnotesize Normal diffusion dependence of the laboratory density $\rho$.
(a): $10^{-2}D_{\perp}$ at $t=14.4$ and $\sim 84\%$ particles left. (b): $10^{-1}D_{\perp}$ at $t=7.2$ and $\sim 67\%$ particles left. (c): $D_{\perp}$ at $t=2.4$ and $\sim 55\%$ particles left. (d): $10D_{\perp}$ at $t=0.36$ and $\sim 46\%$ particles left. The status of total particle number $N_{tot}$ can be understood by comparing figure \ref{fig4}.}
\label{fig19}
\end{figure}

\begin{figure}[h!]
\centering
\includegraphics[scale=0.35]{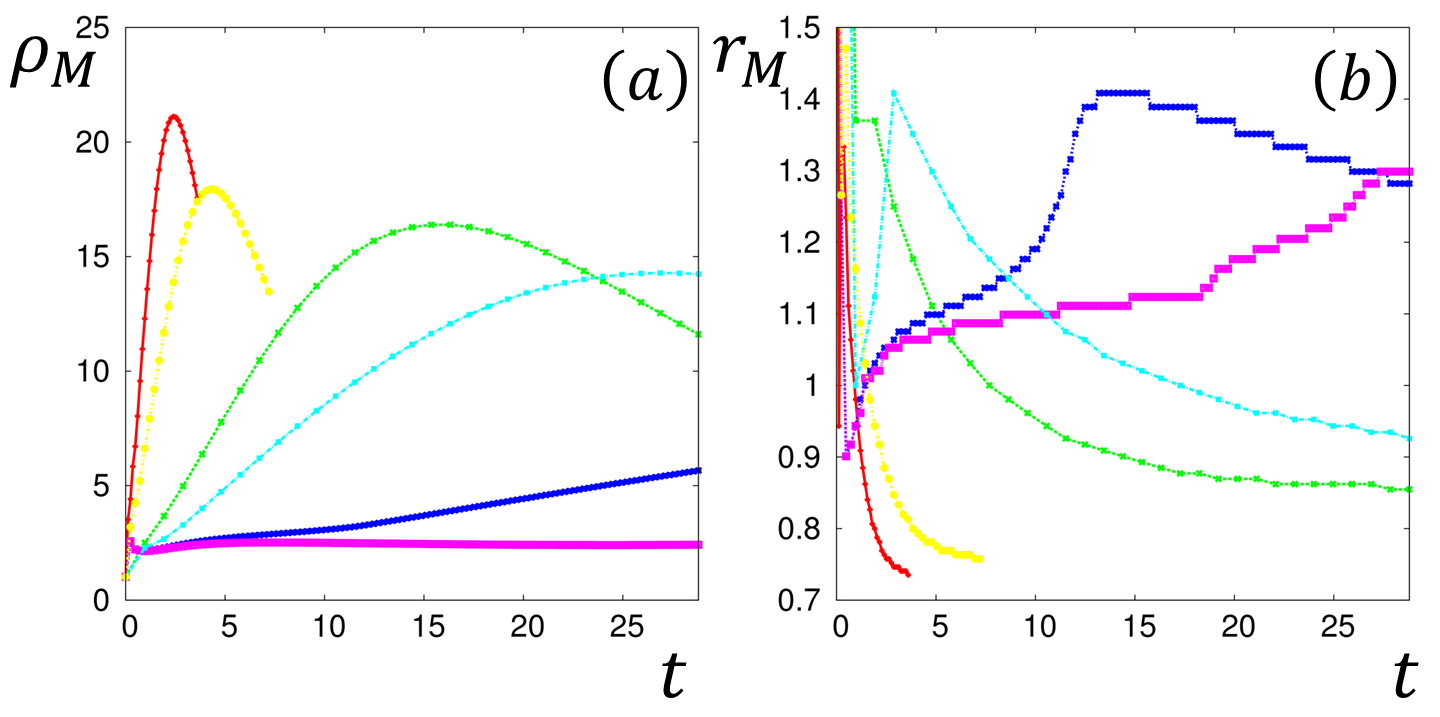}
\caption{\footnotesize (a): time evolution of the instantaneous laboratory density maximum $\rho_{M}$. (b): time evolution of the density maximum radial position $r_{M}$. The color legend is identical to that of figure \ref{fig4}.}
\label{figRM}
\end{figure}

\begin{figure}[h!]
\centering
\includegraphics[scale=0.35]{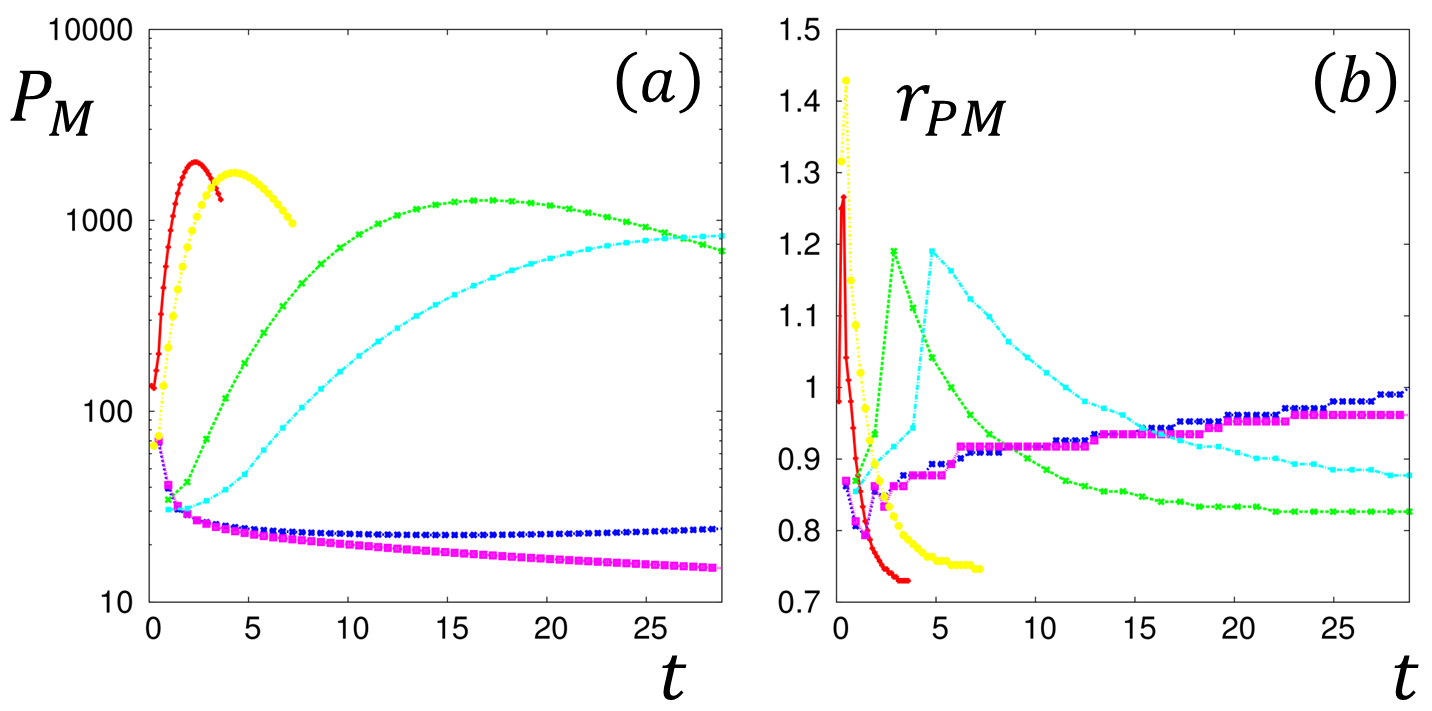}
\caption{\footnotesize (a): time evolution of the instantaneous parallel pressure maximum $P_{M}$. (b): time evolution of the parallel pressure maximum radial position $r_{PM}$. The color legend is identical to that of figure \ref{fig4}.}
\label{figPM}
\end{figure}

\section{Conclusion}

We have derived a model of the inward diffusion
of magnetized particles across flux surfaces in magnetospheric plasmas.
At the core of the formulation is the entropy principle on the symplectic submanifold of the macroscopic hierarchy
that coarse-grains the microscopic action-angle variables (the magnetic moment and the gyro-angle).  
Modeling a random electric field of electrostatic turbulence by a Wiener process,
the diffusion operator $\mathfrak{C}$  
has been derived by careful application of stochastic analysis;
the form of $\mathfrak{C}$ is considerably complicated including
the geometry factor $\mathfrak{q}$ and the phase factor $\alpha$ of stochastic integral.

The geometry factor is due to the anisotropic motion of particles on non-orthogonal coordinates.
When a particle experiences a jump in $\psi$, a simultaneous change in $\ell$ occurs.
This coupling (scaled by $\mathfrak{q}$) yields the diffusion and drift terms in the parallel equation of motion;
the latter is multiplied by $(1/2 -\alpha)$.
Putting $\alpha=0$ (Ito's integral) retains the drift term, while putting $\alpha=1/2$ (Stratonovich's integral)
removes it.
Since the drift term is due to the direct jumps in $\ell$ (not in $\dot{\ell}$), caused by the $E\times B$ drift motion
in $x_\perp$, we conclude that $\alpha=0$ is the right choice
(unlike the usual argument of employing Stratonovich's integral for Brownian motion driven by
jumps in velocities).
Then, the drift term in (\ref{PsiCurr}) pushes particles back to the horizontal plane.

In the Fokker-Planck equation built in Sec. VI, we did not consider a dissipation mechanism acting in the
direction perpendicular to flux-surfaces. 
Hence, particles can be heated by the random electric field,
i.e., the system of particles is an \emph{open system}.
This makes the ``asymptotic form'' of the particle distribution somewhat different from those
predicted by statistical equilibrium models.
For example, the most simple model assumes that the third adiabatic invariant $\psi$ is most fragile, thus
the relaxation will tend $\partial_{\psi}f(t\rightarrow\infty) \rightarrow 0$~\cite{Has}.
Or, assuming a grand canonical ensemble determined by the energy and magnetization, 
a Boltzmann distribution on the foliated (macroscopic) phase space can be derived~\cite{YosFol},
in which $f$ does not depend directly on $\psi$
(however, indirect dependence occurs though $B(\psi,\ell)$ in the Hamiltonian of magnetized particles). 
In the present model, we observe

\begin{equation}
\frac{\partial f}{\partial \psi}\left(t\rightarrow\infty\right)\neq 0\label{nrelaxed} ,
\end{equation}

\noindent which is apparent in (\ref{feq})), as well as in the results of numerical simulations.
Nevertheless, the diffusion term tends to diminish the gradient with respect to $\psi$.

\section{Acknowledgements}

The authors would like to acknowledge the useful advices provided by Y. Kawazura 
with numerical simulations.  

\appendix

\section{Analytical Expression of Point Dipole Field Line Length}

\indent 
We may measure the length $\ell$ of a field line $\tilde{\psi}$ either 
from the inside of the point dipole $(r,z)_{\ell=0}=(0,0)$, 
or from the outside $(r,z)_{\ell=0}=(\{\tilde{r}:\tilde{\psi}=\psi(\tilde{r},0)\},0)$.\\ 
\indent Let us first derive the expression $\ell_{in}(r,z)$ corresponding to the first definition. We start by inverting
the expression for 
$\psi(r,z)$ in order to obtain $z(r,\psi)$:\\

\begin{equation}
z^{2}=r^{\frac{3}{4}}\psi^{-\frac{2}{3}}-r^{2} .\label{eqnA1}
\end{equation}

\noindent Note that we are working with normalized quantities, i.e. $\psi=\psi^{*}/B_{0}L_{0}^{2}$, $z=z^{*}/L_{0}$ and $r=r^{*}/L_{0}$
where $B_{0}$ is the characteristic magnetic field, $L_{0}$ the characteristic length and the $*$ symbol indicates 
physical dimensions. 

Since the stream function $\psi$ is constant along a field line, differentiation gives:

\begin{equation}
\frac{dz}{dr}=\pm\frac{\frac{2}{3}r^{\frac{1}{3}}\psi^{-\frac{2}{3}}-r}{\sqrt{r^{\frac{4}{3}}\psi^{-\frac{2}{3}}-r^{2}}} . \label{eqnA2}
\end{equation}

\noindent The desired length is then:

\begin{widetext}
\begin{equation}
\ell_{in}(r,z)=\int_{0}^{r}{d\ell}=\int_{0}^{r}{\sqrt{\left(1+\left(\frac{dz}{dr}\right)^{2}\right)}dr}=
\int_{0}^{r}{\sqrt{\frac{\frac{4}{9}\psi^{-\frac{4}{3}}-\frac{1}{3}r^{\frac{2}{3}}\psi^{-\frac{2}{3}}}{
r^{\frac{2}{3}}\psi^{-\frac{2}{3}}-r^{\frac{4}{3}}}}dr} .\label{eqnA3}
\end{equation}
\end{widetext}

\noindent To evaluate this integral we perform the change of variable $x=(r\psi)^{\frac{2}{3}}$. This gives:

\begin{widetext}
\begin{dmath}
\ell_{in}(r,z)=\frac{1}{2\psi}\int_{0}^{(r\psi)^{\frac{2}{3}}}{\sqrt{\frac{4-3x}{1-x}}}dx 
=\frac{1}{2\psi}\int_{0}^{(r\psi)^{\frac{2}{3}}}{\frac{1+4-3x+3-3x}{2\sqrt{1-x}\sqrt{4-3x}}}dx 
=\frac{1}{2\psi}\int_{0}^{(r\psi)^{\frac{2}{3}}}{\left(
\frac{1}{2\sqrt{1-x}\sqrt{4-3x}}+\frac{\sqrt{4-3x}}{2\sqrt{1-x}}+
\frac{3\sqrt{1-x}}{2\sqrt{4-3x}}\right)}dx 
=-\frac{1}{2\psi}{\left\{{\frac{\ln{\left[{3\sqrt{1-x}+\sqrt{12-9x}}\right]}}{\sqrt{3}}
+\sqrt{1-x}\sqrt{4-3x}}\right\}}_{0}^{(r\psi)^{2/3}} 
=\frac{1}{2\psi}\left[2-\frac{1}{\sqrt{3}}\ln\left(\frac{\sqrt{3}\sqrt{1-(r\psi)^{\frac{2}{3}}}
+\sqrt{4-3(r\psi)^{\frac{2}{3}}}}{2+\sqrt{3}}\right)
-\sqrt{1-(r\psi)^{\frac{2}{3}}}\sqrt{4-3(r\psi)^{\frac{2}{3}}}\right] .\label{eqnA4}
\end{dmath}
\end{widetext}

\noindent Here $\psi=\psi(r,z)=\frac{r^2}{(r^2+z^2)^{3/2}}$. In a similar manner, it can be shown that the expression for the field line length as measured from the outside of the dipole is:

\begin{widetext}
\begin{equation}
\ell_{out}(r,z)=\int_{\tilde{r}}^{r}{d\ell}=\frac{1}{2\psi}\left[\frac{1}{\sqrt{3}}\ln\left(
\sqrt{3}\sqrt{1-(r\psi)^{\frac{2}{3}}}+\sqrt{4-3(r\psi)^{\frac{2}{3}}}\right)\\
+\sqrt{1-(r\psi)^{\frac{2}{3}}}\sqrt{4-3(r\psi)^{\frac{2}{3}}}\right] .\label{eqnA5}
\end{equation}
\end{widetext}

\section{Change of Variables (Ito's Lemma)}

\begin{THM2}

Define the stochastic integral of a real-valued function $g$ as

\begin{equation}
\int_{s}^{r}gdW=ms\text{-}\lim_{n \to \infty}\sum_{i=1}^{n}g(t_{i-1}+\alpha\Delta t_{i})[W(t_{i})-W(t_{i-1})] ,\label{SIdef}
\end{equation}

\noindent with $\alpha\in[0,1]$. Let $X(\cdot)$ be a real-valued stochastic process satisfying

\begin{equation}
X(r)=X(s)+\int_{s}^{r}Fdt+\int_{s}^{r}G_{\alpha}dW ,
\end{equation}

\noindent with $F\in L^{1}(0,T)$, $G\in L^{2}(0,T)$ and $0\leq s\leq r\leq T$. Consider a new stochastic process $Y=y(X(t),t)$,
where $y:\mathbb{R}\times[0,T]\rightarrow\mathbb{R}$ and $y\in C^{2}$ on its domain. Then $Y$ has the stochastic differential: 

\begin{dmath}
dY(X,t)=\frac{\partial y}{\partial t}dt+\frac{\partial y}{\partial x}dX+\left(\frac{1}{2}-\alpha\right)G^{2}
\frac{\partial^{2} y}{\partial x^{2}}dt=\left(\frac{\partial y}{\partial t}+F\frac{\partial y}{\partial x}+
\left(\frac{1}{2}-\alpha\right)G^{2}\frac{\partial^{2}y}{\partial x^{2}}\right)dt
+G_{\alpha}\left(\frac{\partial y}{\partial x}\right)_{\alpha}dW ,\label{GIL2}
\end{dmath}

\noindent where, as usual, the subscript $\alpha$ indicates evaluation at $t_{\alpha}=t_{i-1}+\alpha\Delta t_{i}$.\\

\end{THM2}

Ito's lemma can be obtained by putting $\alpha=0$. Stratonovich's definition comes instead from the choice $\alpha=1/2$.
It is interesting to see that in this case the stochastic chain rule has the same form of the chain rule
of classic calculus. Physically this is because the Stratonovich interpretation, as we showed in chapter $4$,
describes processes with a continuous variation in time. 

A rigorous proof of the theorem can be obtained by following that given in \cite{Evans} for the case $\alpha=0$.
For the seek of simplicity we give below an informal proof based on Taylor expansion.\\

\textbf{\textit{Proof}}\\

We proceed by Taylor expansion around $t_{i-1}$ and retain only terms of order $dt$:

\begin{widetext}
\begin{dmath}
dY(X,t)=\frac{\partial y}{\partial t}dt+\frac{\partial y}{\partial x}dX+\frac{1}{2}\frac{\partial^{2}y}{\partial x^{2}}dX^{2}+...=\frac{\partial y}{\partial t}dt+\frac{\partial y}{\partial x}Fdt+\left(\frac{\partial y}{\partial x}\right)_{i-1}G_{\alpha}dW+\frac{1}{2}\left(\frac{\partial^{2}y}{\partial x^{2}}\right)_{i-1}G^{2}_{\alpha}dW^{2}+...=
\frac{\partial y}{\partial t}dt+\frac{\partial y}{\partial x}Fdt+\left(\left(\frac{\partial y}{\partial x}\right)_{\alpha}-\left(\frac{\partial^{2}y}{\partial x^{2}}\right)_{\alpha}(X_{\alpha}-X_{i-1})\right)G_{\alpha}dW+\frac{1}{2}\left(\frac{\partial^{2}y}{\partial x^{2}}\right)_{\alpha}G^{2}_{\alpha}dW^{2}+...=
\frac{\partial y}{\partial t}dt+\frac{\partial y}{\partial x}Fdt+\left(\frac{\partial y}{\partial x}\right)_{\alpha}G_{\alpha}dW+\left(\frac{\partial^{2}y}{\partial x^{2}}\right)_{\alpha}\left(\frac{dW^{2}}{2}-dW_{\alpha}dW\right)G^{2}_{\alpha}+...=
\left(\frac{\partial y}{\partial t}+F\frac{\partial y}{\partial x}+\left(\frac{1}{2}-\alpha\right)G^{2}\frac{\partial^{2}y}{\partial x^{2}}\right)dt+\left(\frac{\partial y}{\partial x}\right)_{\alpha}G_{\alpha}dW+O(dt^{3/2}) .
\end{dmath}
\end{widetext}

\noindent Here we used the fact that

\begin{equation}
\langle dW_{\alpha}dW\rangle=\langle(W_{\alpha}-W_{i-1})(W_{i}-W_{i-1})\rangle=\alpha\Delta t_{i} ,
\end{equation}

\noindent and deduced the heuristic formula $dW_{\alpha}dW\sim\alpha dt$. $\blacksquare$\\

\section{Stochastic Integral}

\begin{THM1}\label{THM1}

Define the stochastic integral of a real-valued function $g$ as

\begin{equation}
\int_{s}^{r}gdW=ms\text{-}\lim_{n \to \infty}\sum_{i=1}^{n}g(t_{i-1}+\alpha\Delta t_{i})[W(t_{i})-W(t_{i-1})] ,
\end{equation}

\noindent with $\alpha\in[0,1]$. Let $f=f(X,t)$ be a real-valued function of a random process $X(\cdot)$ satisfying

\begin{equation}
X(r)=X(s)+\int_{s}^{r}Fdt+\int_{s}^{r}GdW ,
\end{equation}

\noindent where the integral is in the Ito sense $\alpha=0$, 
$F\in L^{1}(0,T)$, $G\in L^{2}(0,T)$ and $0\leq s\leq r\leq T$.
Then, the following relation holds:

\begin{equation}
(df)_{Ito}=(df)_{\alpha}-\alpha\frac{\partial f}{\partial x}\frac{\partial G}{\partial x}Gdt .\label{CoI} 
\end{equation}\\

\end{THM1}

What theorem $1$ tells us is that integrating $df$ with Ito's definition $\alpha=0$ is
equivalent to integrating $df$ with a generic $\alpha$ definition, provided that we add
the correction factor $-\alpha\frac{\partial f}{\partial x}\frac{\partial G}{\partial x}Gdt$.\\

\textbf{\textit{Proof}}\\

The first step will be to find the expression of a stochastic differential
equation in the Ito sense in terms of a stochastic differential equation equipped with a
generic $\alpha$ definition of the stochastic integral. 
To accomplish this, consider the Ito SDE

\begin{equation}
dX_{I}=Fdt+G_{i-1}dW ,\label{ISDE}
\end{equation}
 
\noindent where the subscript $I$ indicates Ito's definition $\alpha=0$ of the stochastic integral
and $i-1$ the fact that the function $G$ is evaluated at the left point $t_{i-1}$ of the time interval $\Delta t_{i}=
t_{i}-t_{i-1}$. If we expand $G_{i-1}$ at first order in $dt$ around 
$t_{\alpha}=t_{i-1}+\alpha\Delta t_{i}$
with $\alpha\in[0,1]$, we have 

\begin{dmath}
dX_{I}= Fdt+\left(G_{\alpha}+\left(\frac{\partial G}{\partial x}\right)_{\alpha}(X_{i-1}-X_{\alpha})-
\alpha\left(\frac{\partial G}{\partial t}\right)_{\alpha}\Delta t_{i}\right)dW+...=
Fdt+\left(G_{\alpha}-\left(\frac{\partial G}{\partial x}\right)_{\alpha}(\alpha F \Delta t_{i}+G_{i-1}dW_{\alpha})\right)dW
+...= Fdt+G_{\alpha}dW-\left(\frac{\partial G}{\partial x}\right)_{\alpha}G_{\alpha}dW_{\alpha}dW+...\quad.
\end{dmath}

\noindent Here the subscript $\alpha$ indicates evaluation at point $t_{\alpha}$. Now note that

\begin{equation}
\langle dW_{\alpha}dW\rangle=\langle(W_{\alpha}-W_{i-1})(W_{i}-W_{i-1})\rangle=\alpha\Delta t_{i} .
\end{equation}

\noindent We conclude that

\begin{equation}
dX_{I}=\left[F-\alpha\left(\frac{\partial G}{\partial x}\right)G\right]dt+G_{\alpha}dW\label{Ialpha}+O(dt^{3/2}) .
\end{equation}

\noindent Note that for the terms proportional to $dt$ the choice of $\alpha$ is irrelevant since
they correspond to Riemann integration. Equation (\ref{Ialpha}) is the stochastic differential equation 
in terms of a generic $\alpha$-integral 
corresponding to the stochastic differential equation in the Ito sense (\ref{ISDE}). 

We are now ready to prove the main result. Using again Taylor expansion around $t_{\alpha}$ and
retaining only terms of order $dt$, we have

\begin{widetext}
\begin{dmath}
(df)_{I}=\frac{\partial f}{\partial t}dt+\frac{\partial f}{\partial x}dX_{I}+
\frac{1}{2}\frac{\partial^{2} f}{\partial x^{2}}dX_{I}^{2}+...=
\frac{\partial f}{\partial t}dt+\frac{\partial f}{\partial x}Fdt+
\left(\frac{\partial f}{\partial x}\right)_{i-1}G_{i-1}dW+
\frac{1}{2}\frac{\partial^{2} f}{\partial x^{2}}(G_{i-1}^{2}dW^{2})+...=
\left(\frac{\partial f}{\partial t}+\frac{\partial f}{\partial x}F+\frac{1}{2}\frac{\partial^{2} f}{\partial x^{2}}
G^{2}\right)dt+\left(\left(\frac{\partial f}{\partial x}\right)_{\alpha}+\left(\frac{\partial^{2} f}{\partial x^{2}}\right)_{\alpha}(X_{i-1}-X_{\alpha})\right)\left(G_{\alpha}+\left(\frac{\partial G}{\partial x}\right)_{\alpha}
(X_{i-1}-X_{\alpha})\right)dW+...=
\left(\frac{\partial f}{\partial t}+\frac{\partial f}{\partial x}F+\frac{1}{2}\frac{\partial^{2} f}{\partial x^{2}}
G^{2}\right)dt+\left(\frac{\partial f}{\partial x}\right)_{\alpha}G_{\alpha}dW-\left(\frac{\partial f}{\partial x}\right)_{\alpha}\left(\frac{\partial G}{\partial x}\right)_{\alpha}G_{\alpha}dW_{\alpha}dW-
\left(\frac{\partial^{2} f}{\partial x^{2}}\right)_{\alpha}G^{2}_{\alpha}dW_{\alpha}dW+...=
\left(\frac{\partial f}{\partial t}+\frac{\partial f}{\partial x}F+\left(\frac{1}{2}-\alpha\right)\frac{\partial^{2} f}{\partial x^{2}}
G^{2}-\alpha\frac{\partial f}{\partial x}\frac{\partial G}{\partial x}G\right)dt+\left(\frac{\partial f}{\partial x}\right)_{\alpha}G_{\alpha}dW
+...=
(df)_{\alpha}-\alpha\frac{\partial f}{\partial x}\frac{\partial G}{\partial x}Gdt+O(dt^{3/2}) .
\end{dmath}
\end{widetext}

\noindent Here we used the generalized Ito's lemma (\ref{GIL2}) obtained in appendix B to identify $(df)_{\alpha}$. $\blacksquare$

Note that when $G$ is constant $(df)_{I}=(df)_{\alpha}$.

\section{The $\ell$-Equation}

Since $\ell$ is a function of $x_{\perp}$, when evaluating how $\ell$ is affected by a random change in $x_{\perp}$ Ito's lemma (\ref{GIL2}) has to be applied accordingly. Recalling (\ref{stocvper}), we have

\begin{widetext}
\begin{dmath}
(dL)_{turbulence}=\left(\frac{1}{2}-\alpha\right)\frac{D_{\perp}}{(rB)^{2}}\frac{\partial^{2}\ell}{\partial x_{\perp}^{2}}dt+\mathfrak{q}D_{\perp}^{1/2}dW_{\perp}=
\left(\frac{1}{2}-\alpha\right)\frac{D_{\perp}}{(rB)^{2}}\frac{\partial}{\partial x_{\perp}}\left(\mathfrak{q}rB\right)+\mathfrak{q}D_{\perp}^{1/2}dW_{\perp}=
\left(\frac{1}{2}-\alpha\right)\frac{D_{\perp}}{rB}\left(\mathfrak{q}\frac{\partial}{\partial \ell}+\frac{\partial}{\partial\psi}\right)\left(\mathfrak{q}rB\right)+\mathfrak{q}D_{\perp}^{1/2}dW_{\perp}=
\left(\frac{1}{2}-\alpha\right)D_{\perp}\left[\left(\mathfrak{q}\frac{\partial}{\partial \ell}+\frac{\partial}{\partial\psi}\right)\mathfrak{q}+\mathfrak{q}\left(\mathfrak{q}\frac{\partial}{\partial \ell}+\frac{\partial}{\partial\psi}\right)\ln(rB)\right]dt+\mathfrak{q}D_{\perp}^{1/2}dW_{\perp} ,
\end{dmath}
\end{widetext}

\noindent where the following relationship was used:

\begin{equation}
\begin{split}
\partial_{\perp}&=\frac{\partial}{\partial x_{\perp}}=\frac{\partial \ell}{\partial x_{\perp}}\frac{\partial}{\partial \ell}+\frac{\partial\psi}{\partial x_{\perp}}\frac{\partial}{\partial\psi}\\=&\nabla \ell\cdot \partial_{\perp}\partial_{\ell}+rB\partial_{\psi}=rB\left(\mathfrak{q}\partial_{\ell}+\partial_{\psi}\right) .
\end{split}
\end{equation}

\noindent Taking into account $v_{\parallel}$ gives (\ref{l}).

\end{document}